\documentclass{revtex4}

\usepackage{graphicx}
\usepackage{dcolumn}
\usepackage{bm}
\usepackage{epstopdf}
\usepackage{multirow} 
\begin{document}
\preprint{APS/123-QED}

\title{Local correlation method for metals: benchmarks for surface and adsorption energies}

\author{Elena Voloshina\footnote{E-mail: elena.voloshina@fu-berlin.de} }
\affiliation{\mbox{Physikalische und Theoretische Chemie, Freie Universit\"at Berlin, 14195 Berlin, Germany}
}

\date{\today}

\begin{abstract}

Highly accurate methods such as coupled cluster (CC) techniques can be used for periodic systems within the framework of the method of increments. Its extension to low-dimensional conducting system is considered. To demonstrate the presented approach a clean Mg (0001) surface is selected, where the CC treatment with single and double excitations and perturbative triples is used for calculation of the surface energy. Further example concerns the adsorption energy of Xe on the metal surface. The obtained results can be used to verify performance of the approximate methods. Along with  the computational speed-up at the high-level of accuracy, application of the method of increments provides for a possibility to analyze influence of individual correlation-energy increments on the studied property. 
 
\end{abstract}

\maketitle

\section{Introduction}

Although density functional theory (DFT) remains the most frequently used method in the field of solid-state physics and material science, there is no clear way to systematically improve results obtained with the present-day density functionals. This yields preferences to the wavefunction-based correlation techniques, a hierarchy of methods that, starting from the one-electron Hartree-Fock (HF) solution, allows one to approximate the many-body nature of the true ground-state wavefunction in a rigorous and systematic manner. One of the largest problems standing on the way of the application of such methods to extended systems is their unfavorable scaling with respect to the increasing system size. From a physical point of view, however, this difficulty should be avoidable because the correlation hole around an electron is a fairly local object. 

The reformulation of the many-body wavefunction in terms of localized orbitals defines the group of the so-called local correlation methods (see e.g. Refs.~\cite{lcm1,lcm2,lcm3,lcm4,lcm5,lcm6,method}). One method of this type, the method of increments~\cite{method}, combines HF calculations for periodic systems (which give Coulomb and exchange terms of the total energy) with correlation calculations for finite embedded fragments of a solid, and the total correlation energy is written as a cumulant expansion in units of contributions from localized orbital groups of increasing size:
\begin{equation}
\label{corr}
    E^{\mathrm{corr}}=\sum_{i}\varepsilon_{i}+\frac{1}{2!}\sum_{i\ne j}\Delta\varepsilon_{ij}+
    \frac{1}{3!}\sum_{i\ne j \ne k}\Delta\varepsilon_{ijk} + \ldots \, .
\end{equation}
The summation over $i$ includes all orbitals in the reference cell, while $j$ and $k$ extend over the entire system. The one-body increment ($\varepsilon_i$) involves all excitations from the orbitals $i$, while the two-body increment ($\Delta\varepsilon_{ij}$) includes the nonadditive excitations summed over orbitals $i$ and $j$. In order to get the latter quantity, one subtracts from the correlation energy, obtained for the joint orbital system $ij$, the corresponding one-body contributions: $\Delta\varepsilon_{ij}=\varepsilon_{ij}-(\varepsilon_{i}+\varepsilon_{j})$. Higher-order increments are defined in analogous way, e.g.\,for the three-body term we get: $\Delta\varepsilon_{ijk}=\varepsilon_{ijk}-(\varepsilon_{i}+\varepsilon_{j}+\varepsilon_{k})-(\Delta\varepsilon_{ij}+\Delta\varepsilon_{ik}+\Delta\varepsilon_{jk})$. In the limit of the sum over all $i$, $j$, $k$ as well as over orders of increments, one would obtain the exact correlation energy per unit cell of the infinite system. In practice, the neglect of four-body and higher-order contributions is possible and the incremental expansion converges swiftly with distance (see Ref.~\cite{beate-habil} and the references therein). The translation symmetry of crystals reduces the amount of increments to be evaluated drastically. In order to use this advantage, one determines the so-called weight factors ($\omega$) with which the individual increments occur in the considered structure. 

The presence of well-localized orbitals is a precondition for the applicability of local correlation schemes. For metals, it is no longer possible to generate well-localized Wannier orbitals by unitary transformation within the occupied HF space. To overcome this difficulty it has been suggested to base the incremental expansion on a well-localizable model system rather than on the real metal and to allow for delocalization (and thereby approaching the real metal) only gradually within the various levels of the many-body expansion~\cite{embed}. This scheme is tested for series of bulk metals and the obtained cohesive properties are shown to be in very good agreement with experimental results~\cite{hg1,mg1,zn1,cd,met-review}.

Due to the need to calculate the correlation energy increments in finite fragments of the solid, one has to construct an embedding, which simulates the conditions in the infinite system. The embedding is chosen in such a way that the localized orbitals in the finite systems are quite similar to the Wannier orbitals of the periodic system. Type of the embedding depends on the nature of binding in the system under consideration, e.g. dangling bonds saturation with hydrogen atoms in the case of covalent systems (see e.g. Refs.~\cite{si,h2o}) or a point charge array in the case of an ionic crystal (see e.g. Refs.~\cite{ceo2,carsten}). When dealing with metal, the solution is not obvious  since the correlation effects that influence the ground-state properties are quite local in this case. However, an attempt to get good agreement with experiment for Mg-bulk correlation energy using the free clusters fails due to the mixing of species of different nature~\cite{mg1}. That is way different variants of the embedding and the localized orbitals used for the incremental scheme have been developed and tested; their advantages and disadvantages are discussed in Refs.~\cite{embed,met-review}.

Being defined for three-dimensional case, the method of increments can nevertheless be applied for systems with reduced symmetry. For instance, to quantify the molecule-surface interaction the adsorption energy can be written as $E_\mathrm{ads}=E_{S+A}-E_{S}-E_{A}=E_\mathrm{ads}^\mathrm{HF}+E_\mathrm{ads}^\mathrm{corr}$, where $E_{S+A}$ is the total energy of the substrate/adsorbate system, and $E_S$ and $E_A$ are the energies of the fragments at the same coordinates as in the substrate/adsorbate system corrected for the basis set superposition error. Again, the HF energy, $E_\mathrm{ads}^\mathrm{HF}$, is calculated for the periodic system in the standard way, and the correlation contribution, $E_\mathrm{ads}^\mathrm{corr}$, is calculated within the incremental expansion:
\begin{equation}
\label{ads}
E^\mathrm{corr}_\mathrm{ads}=\eta_A + \sum_i\eta _{i}  + \sum_{i}\eta _{Ai} + \sum_{i<j}\eta _{ij} + \sum_{i<j}\eta _{Aij}+  \ldots \, .
\end{equation}
Here: $\eta_A$ is the correlation contribution to the adsorption energy within the adsorbate, $\eta_i$ and $\eta_{ij}$ are the changes in the surface increments due to adsorption, $\eta_{Ai}$ are caused by the interaction between localized orbitals of the adsorbate and the surface. Overall, one has to take into account all orbital groups, that change due to the interaction of an adsorbate with a substrate.  This approach can be especially efficient in the cases where DFT is less successful, e.g. when dealing with phy\-si\-sorp\-tion~\cite{carsten,h2s,h2o}, although there is no doubt that chemisorption can be investigated as well.

Combining the incremental approach with the scheme proposed for metals, it should be possible to consider metallic surfaces. In this manuscript, we present the results on the Mg(0001) surface and the Xe/Mg(0001) system, where the above approach is used to investigate surface stability as well as interaction energy between the metal surface and the adatom. 

\section{\label{sec:methods}Computational details}

The periodic mean-field calculations for the studied system were performed with the program package CRYSTAL\,09~\cite{crystal}. In order to obtain converged results for the HF binding energy, the default parameters were modified. For the pseudo-overlap tolerances ITOL$4$ and ITOL$5$, used for the prescreening of the exchange integrals, tighter values than usual have been employed: $15$ and $80$, respectively. The other CRYSTAL tolerances (ITOL$1$-$3$), used for the screening of Coulomb integrals, have been set to $7$. The shrinking factors defining the $k$-mesh were set to $16$ and $32$. The basis sets employed are the crystal-optimized all-electron basis set of valence-double-$\zeta$ (VDZ) quality including polarization functions~\cite{mg-wedig} for Mg and the one optimized for the $8$-valence electron pseudopotential~\cite{xe-basis1} for the Xe atom. The adsorbate concentration is set to approximately $3.7\cdot 10^{-2}\,\mathrm{\AA^{-2}}$. The converged result for the surface energy at the HF level is achieved for the slab consisting of  feefteen Mg-layers. When calculating adsorption energy, the seven-layer Mg-slab was used.

For compatibility, the DFT calculations reported here were performed with the same thresholds and basis set as used for the HF calculations. For the exchange-correlation energy functional we employed the local density approximation (LDA)~\cite{lda} and the generalized gradient approximation as parameterized by Perdew \textit{et al.} (PBE)~\cite{pbe}. In the context of DFT, the contributions of the long-range van der Waals interactions on the adsorption energy of the xenon on magnesium were accounted for by means of the semi-empirical approach proposed by Grimme~\cite{grimme}. 

\begin{figure}
\includegraphics[width=0.65\textwidth]{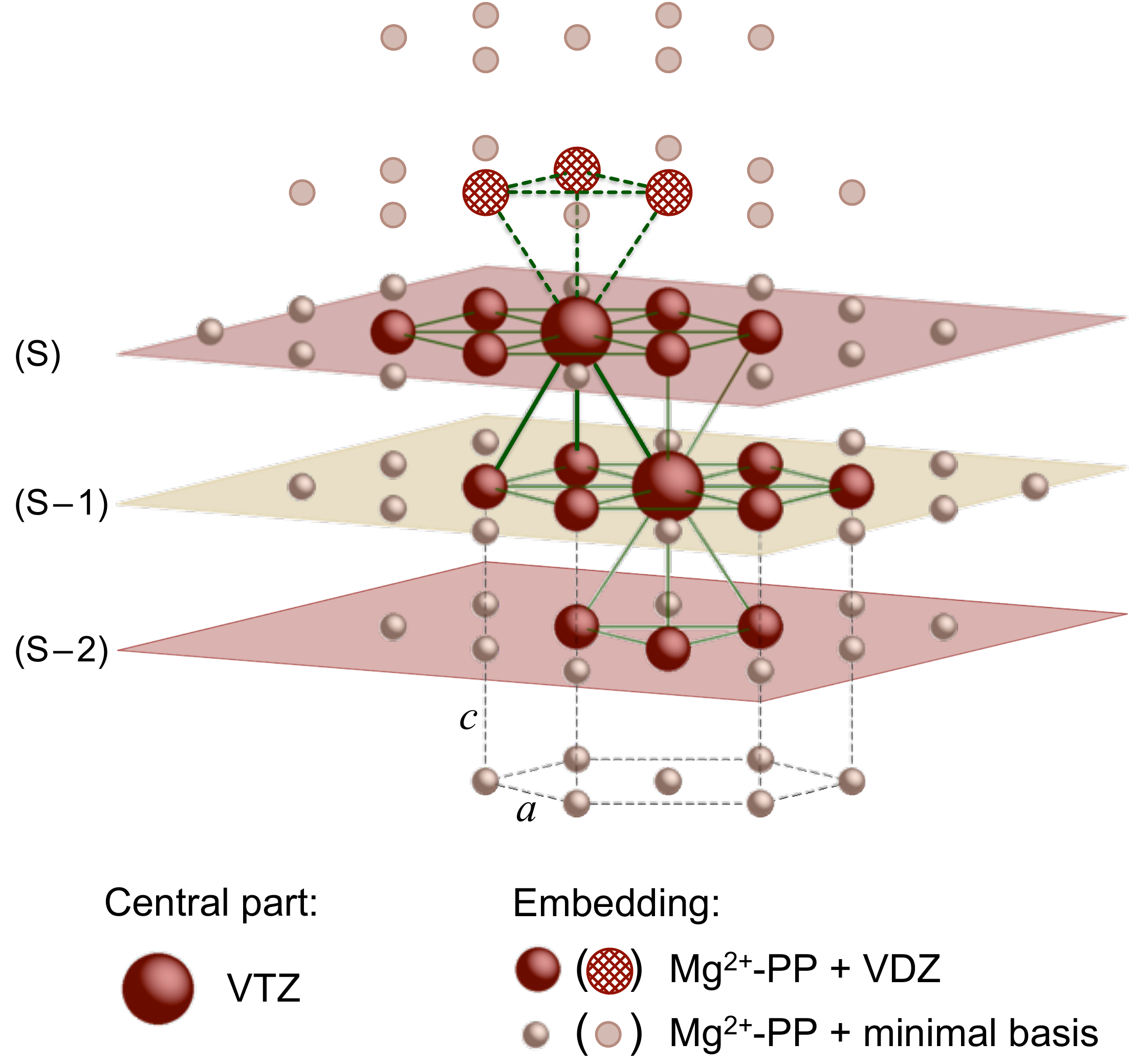}
\caption{\label{fig:cluster} Example of a ``bulk'' Mg-cluster for calculating the 2-body contribution to the correlation energy  ($r_{12}=3.197$\,\AA). Spheres of different size and style represent atoms of different types (\textit{c.f.} text). For the one of the central atoms, six nearest-neighbor distances are depicted with thick lines (either solid or dashed). Atoms shown with plain spheres (either hatched or filled) disappear when constructing the ``surface'' cluster. Crystallographic structure of $hcp$ Mg is shown with thin dashed lines.}
\end{figure}

The correlation-energy increments are calculated for selected fragments which reflect the geometry of the crystal. These fragments have two components, firstly the atoms $i$, $j$, $k$, ... to be correlated (in the center of the fragment) and secondly the embedding atoms. We select the positions of the atoms to be correlated and surround each of these atoms by embedding shell with a cutoff for the embedding of $2$ times the in-plane lattice parameter of the $hcp$ Mg. This results in the cluster consisting of $51$ atoms ($1$ atom in the central part and $50$ embedding atoms), when considering the one-body contribution. To get the results presented in this work the embedding was partitioned into two regions, a true embedding region (second embedding shell), where the atoms are described with a minimal basis, and an intermediate region (first embedding shell), where the occupied orbitals are kept frozen, but additional basis functions are supplied for the re-optimization of the orbitals in the central part of cluster. An example of the embedded two-body cluster is shown in  Fig.~\ref{fig:cluster}.

Correlation calculations are performed with the program package MOLPRO~\cite{molpro}, using the coupled-cluster treatment with single and double excitations and perturbative triples [CCSD(T)]~\cite{ccsdt}. For the perturbative approaches in the incremental scheme, we have re-canonicized the space of the localized orbitals which are correlated to obtain a diagonal Fock matrix. The basis sets employed are the cc-pVTZ of Dunning~\cite{mg-basis2} for Mg and the cc-pVTZ optimized for the $8$-valence electron pseudopotential~\cite{xe-basis2} increased by even-tempered basis functions to $[4s4p3d2f]$ for the Xe atom. Embedding atoms are described with the $2$-valence electron pseudopotential~\cite{mg-basis3} and either the corresponding basis set of VDZ quality (first embedding shell, see Fig.~\ref{fig:cluster}) or a minimal basis set (second embedding shell, see Fig.~\ref{fig:cluster}).

\section{Results and discussion}

Surface energy ($E_\mathrm{surf}$) quantifies the disruption of bonds that occurs when a surface is created. Within the slab approach, mathematically this can be written as follows: $E_\mathrm{surf}=(E_\mathrm{slab}-n\times E_\mathrm{bulk})/2$ ($n$ is a number of layers in the slab). Therefore in the formalism of the method of increments, correlation contribution to the surface energy ($\gamma$) can be determined as a difference between the increment calculated using the ``surface'' cluster (Fig.~\ref{fig:cluster}), multiplied with its weight factor ($\omega^\mathrm{surf}$), and the corresponding one calculated using the ``bulk'' cluster, multiplied with its weight factor ($\omega^\mathrm{bulk}$). For example, for the two-body correlation-energy increment one writes:
\begin{equation}
\label{surf}
\small
\gamma_{ij}=\Delta\varepsilon_{ij}^\mathrm{surf}\times\omega_{ij}^\mathrm{surf}-\Delta\varepsilon_{ij}^\mathrm{bulk}\times\omega_{ij}^\mathrm{bulk}.
\end{equation} 

\begin{figure}
\includegraphics[width=0.65\textwidth]{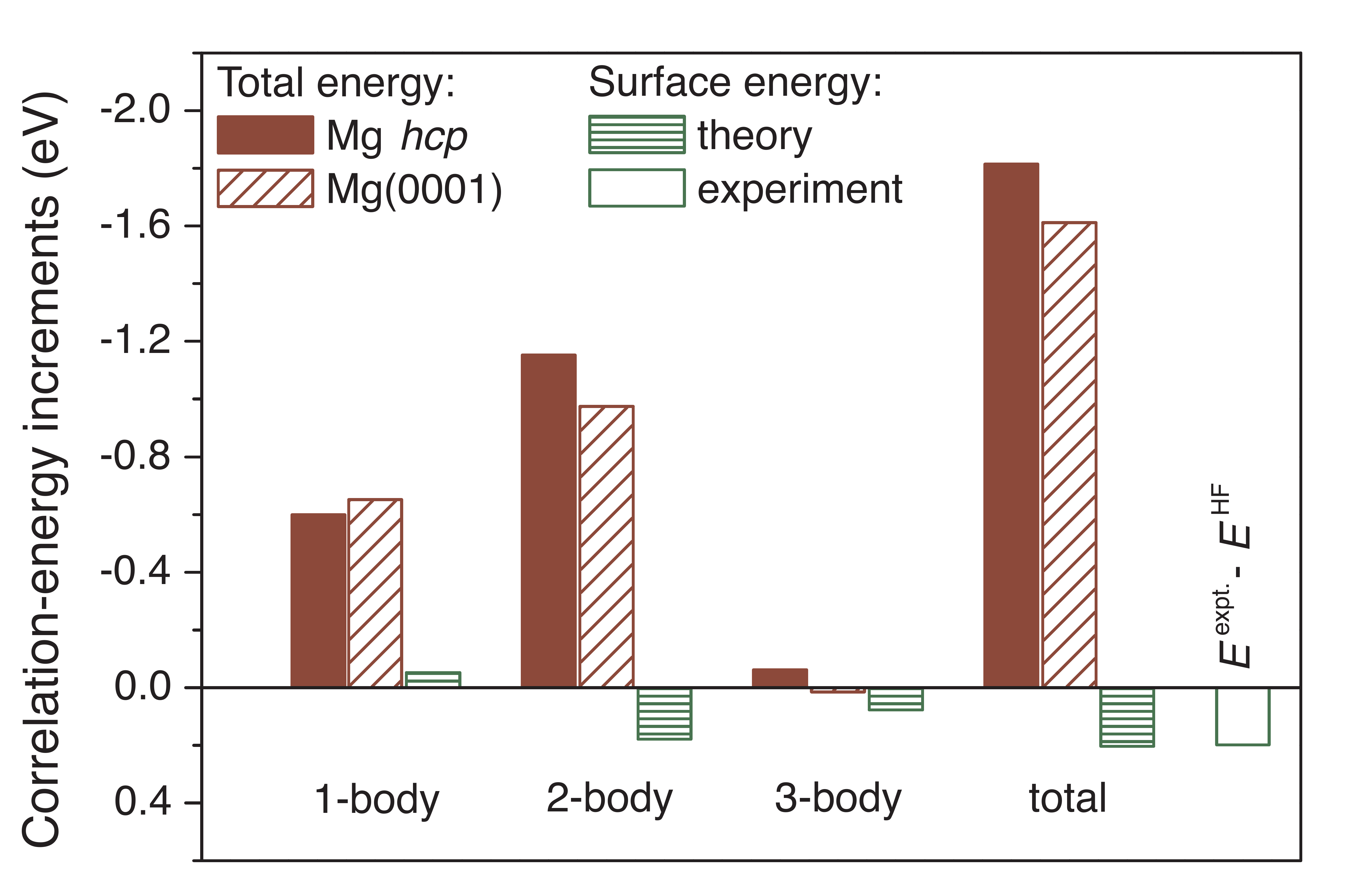}
\caption{\label{fig:bulk+surf} Correlation contributions to the total energy of $1$-, $2$-, and $3$-body increments and calculated surface energy as compared with experimental value.}
\end{figure}

In the physics of solids, surfaces must be intrinsically less energetically favorable than the bulk of a material. From the computational point of view this means positive sign for the value of the surface energy. The experimental result for the Mg(0001) surface is $+0.383$\,eV/atom~\cite{mg_surfen_expt}. HF method, neglecting electron correlation effects, yields only $47$\,\% of $E_\mathrm{surf}$ (Tab.~\ref{tab:tab1}). Summing up 1-, 2-, and 3-body contributions one gets $E_\mathrm{surf}^\mathrm{corr}=+0.20361$\,eV (Fig.~\ref{fig:bulk+surf}), that leads to $E_\mathrm{surf}^\mathrm{total}=+0.385$\,eV (Tab.~\ref{tab:tab1}), that is in very good agreement with the experimental value. 

The adsorption of rare-gas atoms on metal surfaces represents prototypical physical adsorption systems and, thus, Xe adatom on Mg(0001) can be viewed as a good model to test our approach. Two adsorption sites have been investigated: (i) a low-coordinated $atop$ position and (ii) the $fcc$ hollow site (The corresponding dissociation curves are presented in Supplementary Fig. S1). The obtained results are summarized in Tab.~\ref{tab:tab1}. Whereas HF shows no binding in both cases, correlations stabilize the structures, giving preference to the $atop$ geometry. Up to our knowledge, there are no experimental data regarding the studied Xe/Mg(0001) system, however, the $atop$ position was experimentally found to be more favorable, when investigating Xe adatom on Cu(111), Pd(111), Pt(111), and Nb(110)~\cite{xe-met1,xe-met2}. 

It is interesting to compare the results evaluated by means of the presented approach with those of DFT. The data obtained employing LDA
and PBE are listed in Tab.~\ref{tab:tab1}. As usual, LDA overestimates the binding, that leads to too short lattice parameters of the $hcp$-Mg and too large $E_\mathrm{coh}$. Further, surface is c.a.\,$4$\,\% less stable than it is experimentally and $E_\mathrm{ads}$ is strongly overestimated compared to the benchmark. Although PBE results are in reasonable agreement with experiment for bulk characteristics, this functional yields overestimation of the surface stability, demonstrating behavior typical for gradient-corrected functionals~\cite{kresse}. Finally, PBE fails to describe Xe phy\-si\-sorp\-tion, that is not surprising as well. In the latter case, performance of PBE can be improved when employing DFT-D2 scheme of Grimme~\cite{grimme}. This way, adsorption energies and equilibrium distances are in good agreement with the presented CCSD(T) results, deviating by less than $1$\,\%.

An asset of the incremental scheme is that along with the computational speed-up at the high-level of accuracy one gets the possibility to analyze various contributions to the correlation energy. This gives us a chance to penetrate deeper into the nature of binding and answer some interesting questions. The first one is:  why do correlations destabilize the surface? One can see from Fig.~\ref{fig:bulk+surf}, that $E_\mathrm{surf}^\mathrm{corr}$ is mainly determined by the two-body contributions. Therefore, in order to find an answer to this question, it is reasonable to analyze the two-body correlation-energy increments considered in this work (Tab.~\ref{tab:tab2}). Indeed, $\Delta\varepsilon_{ij}^\mathrm{surf}$ are higher in magnitude as compared to the corresponding contributions of bulk. At the same time, in most cases $\gamma_{ij}$ is larger than $0$, that is due to the lower weight factors $\omega_{ij}^\mathrm{surf}$. Weight factor for the two-body increment is defined as the number of neighboring atoms located at the certain distance from the one set as a central (divided by two in order to get the contribution per one atom). In Fig.~\ref{fig:cluster}, where the cluster used for the calculation of $\Delta\varepsilon_{12}$ is presented, the six distances of one of the central Mg in the ``bulk'' cluster are shown with thicker lines; three of them, drawn with dashed lines, will disappear, when constructing the corresponding ``surface'' cluster. As a result, the corresponding weight factor is reduced by $2$. The situation is different, when considering ``surface'' clusters with two atoms lying in the top, (S), layer. In such cases (footnoted with \textit{a} in Tab.~\ref{tab:tab2}), $\omega_{ij}^\mathrm{surf}=\omega_{ij}^\mathrm{bulk}$ leading to negative $\gamma_{ij}$. Summing up all $\gamma$, one gets $E_\mathrm{surf}^\mathrm{corr}$, which is large than $0$.

\begin{figure}
\includegraphics[width=0.50\textwidth]{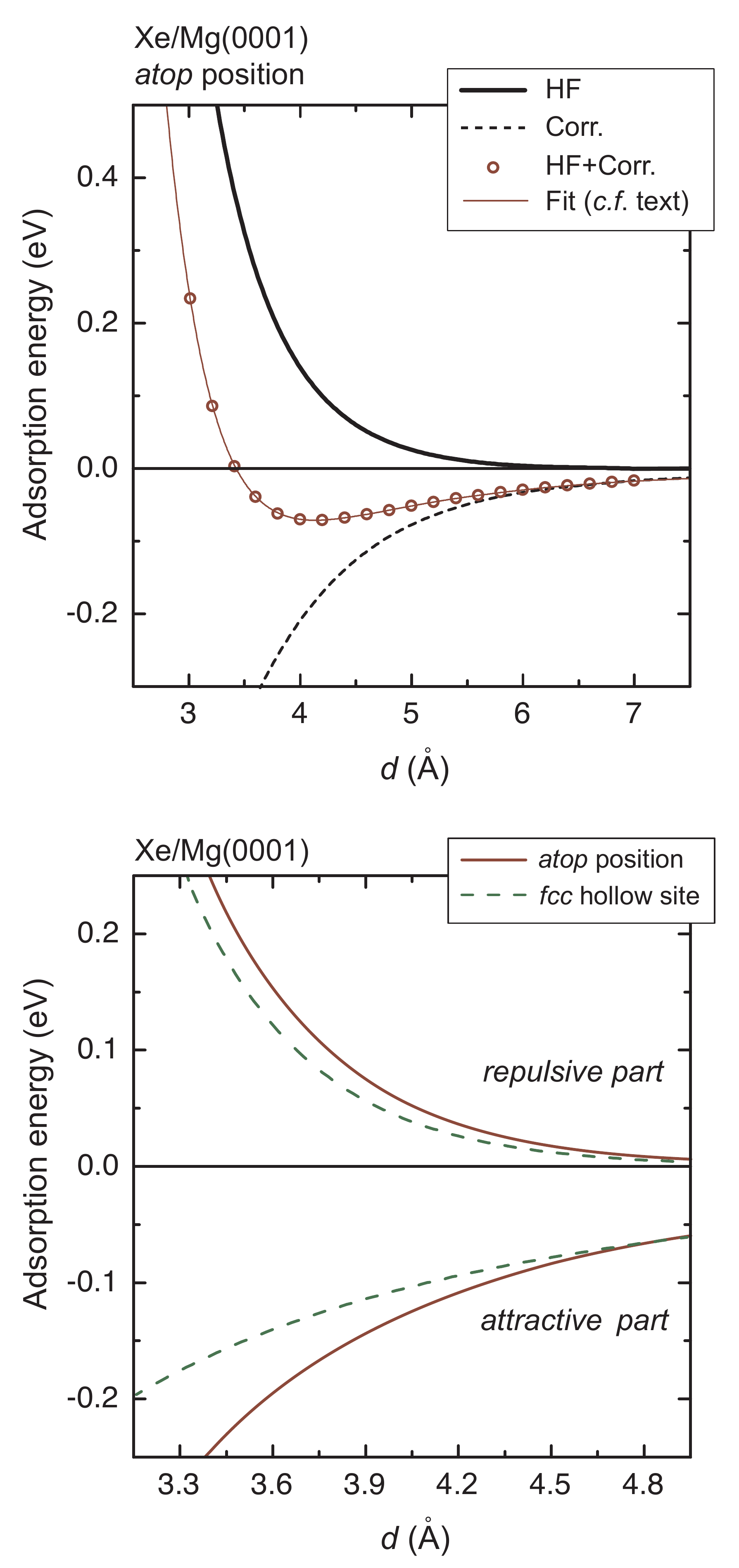}
\caption{\label{fig:analysis1} 
(Upper panel) The different contributions to the adsorption energy as functions of substrate-adsorbate distance are plotted for the Xe/Mg(0001) system as obtained for the $atop$ adsorption geometry. Circles represent the data points, obtained by summing up the HF (thick solid line) and correlation (dashed line) parts of energy. Thin solid line is a fit according to the formula $E(d)=A\,\mathrm{exp}(-Bd)-C/(d-D)^3$ (\textit{c.f.} text).
(Lower panel) Decomposition of the binding energy curve into repulsive and attractive contributions (\textit{c.f.} text). }
\end{figure}

The second question can be formulated as follows: Noting that Xe adatom has closed shell, it can be assumed that this adsorbate should prefer maximally coordinated sites (minimizing the Pauli repulsion). What is the reason for the discrepancy between this expectation and our observations? In order to understand this phenomenon we decomposed the calculated binding-energy curves using an empirical formula as suggested in Ref.~\cite{scheffler}. In a contrast to standard DFT approximations, where asymptotic behavior of the curves cannot be described due to the built in locality~\cite{scheffler}, in the case of the approach presented in this work,  $E(d)=A\,\mathrm{exp}(-Bd)-C/(d-D)^3$ represents the perfect fit for the computed data-points [see Fig.~\ref{fig:analysis1} (Upper panel) as well as Supplementary Fig. S1]. The corresponding energy decomposition into repulsive and attractive potential is shown in Fig.~\ref{fig:analysis1} (Lower panel). One can see, that indeed, Pauli repulsion is weaker, when Xe occupies the $fcc$ hollow site. At the same time, an attractive part of the potential is weaker in this case as well, when compared to that for the $atop$ geometry. 

\begin{figure}
\includegraphics[width=0.65\textwidth]{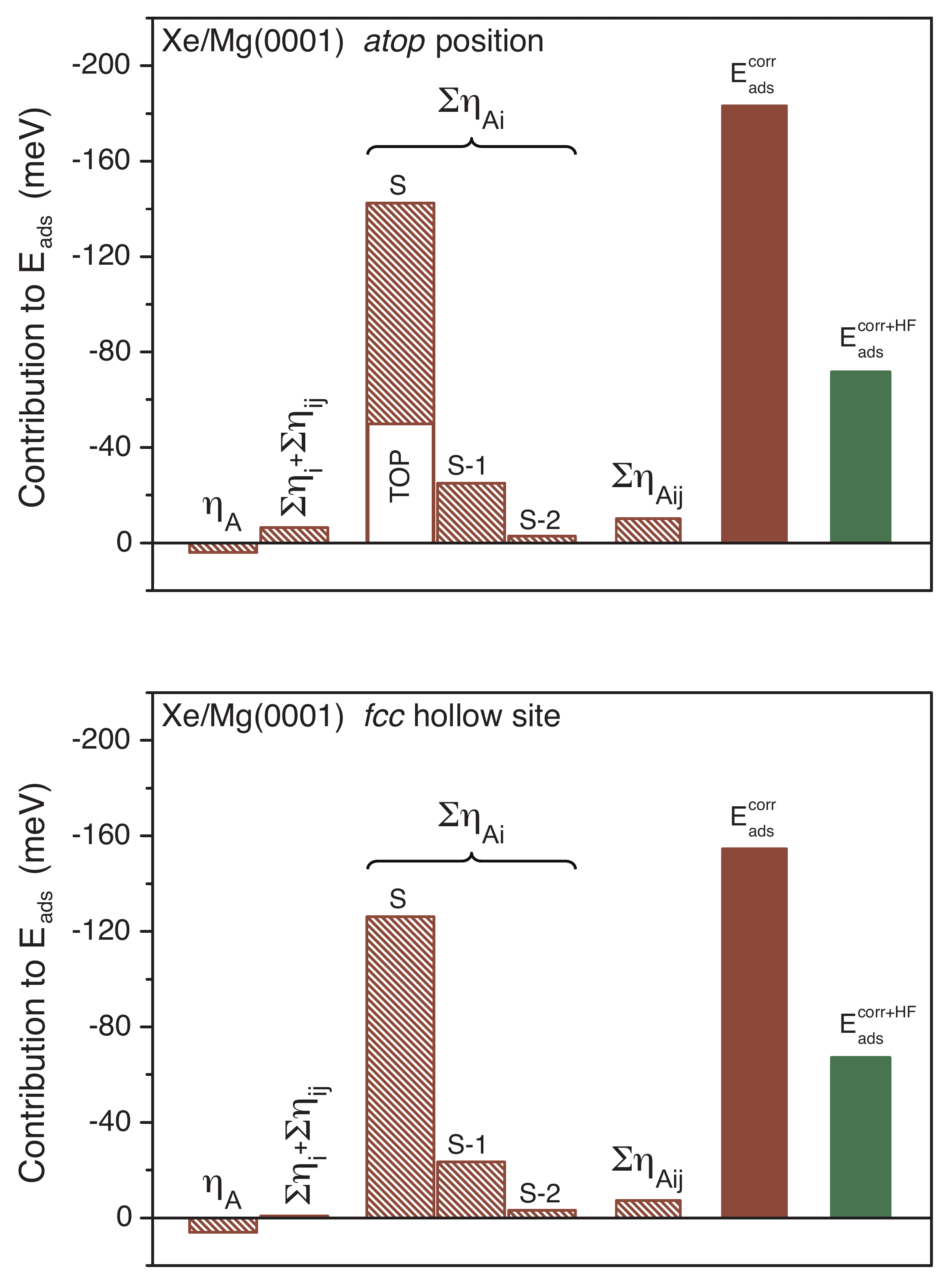}
\caption{\label{fig:mg+xe_top} The different contributions to the total adsorption energy are plotted for the Xe/Mg(0001) system in its equilibrium geometry, when Xe is occupying the $atop$ position (upper panel) or the $fcc$ hollow site (lower panel). }
\end{figure}

\begin{figure}
\includegraphics[width=0.50\textwidth]{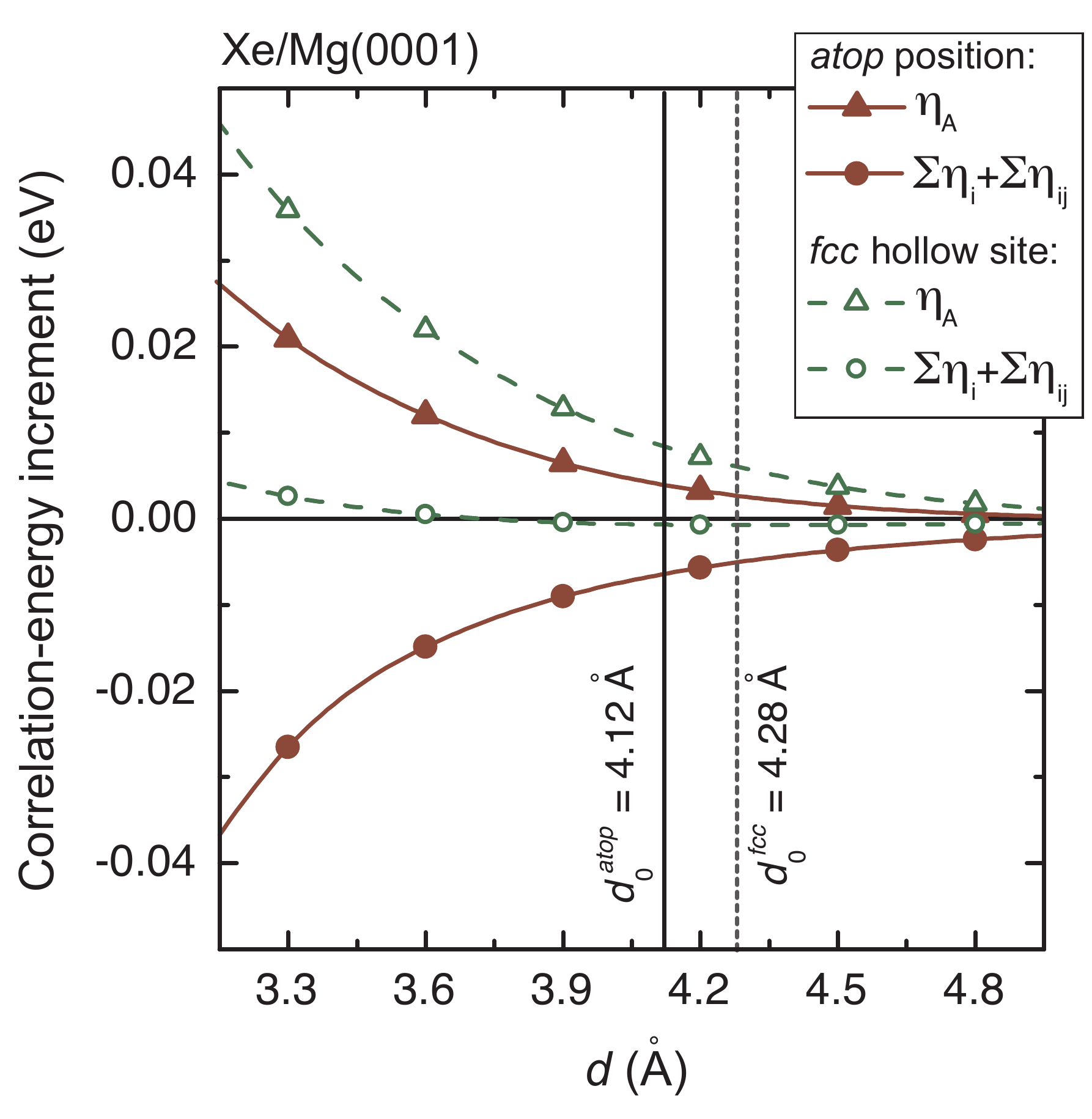}
\caption{\label{fig:analysis2} Correlation contributions to the adsorption within the adsorbate ($\eta_A$) and within the surface ($\sum\eta_i+\sum\eta_{ij}$) as functions of distance between Xe and the magnesium surface. Equilibrium distances obtained for the two adsorption positions are shown with vertical lines.  }
\end{figure}

As a further step in the understanding, we can explore the advantage of the method of increments to analyze the individual contributions to $E_\mathrm{ads}^\mathrm{corr}$ computed according to eqn.~(\ref{ads}). The different contributions to the total adsorption energy as obtained for the two considered situation are presented in Fig.~\ref{fig:mg+xe_top}. As expected, the major contribution to the adsorption energy is from the two-body correlation energy increments $\eta_{Ai}$. Among them, the interactions between Xe atom and interface Mg layer, (S), are the most important, in particular, one third is due to the electron correlations of  the adsorbate and the top Mg atom, when the $atop$ geometry is considered [indicated with ``TOP'' in Fig.~\ref{fig:mg+xe_top} (Upper panel)].  $\eta_{Ai}$ involving electrons of Mg-atoms of lower lying layers, (S-1) and (S-2), contribute to the $\sum\eta_{Ai}$ with ca. $15$\,\% and ca. $2$\,\%, respectively. One may note, that the major difference between the two adsorption geometries is observed for the increments involving orbitals of entire adsorbate or entire substrate. Their dependance from the Xe-Mg(0001) distance is plotted in Fig.~\ref{fig:analysis2}. The curve related to the polarization of the Xe atom is more attractive for the $atop$ geometry as compared to the case when adatom is occupying the $fcc$ hollow site, indicating the higher dipole moment of Xe induced by its interaction with metal surface. Still, the $atop$- and $fcc$-related curves show similar behavior. This is not the case, when analyzing the results obtained for the sum of the one- and two-body increments involving the orbitals of Mg-surface. Whereas the function obtained for the $atop$ geometry is attractive for all considered distances, in the case when Xe is occupying the $fcc$ hollow site it is weakly repulsive for short distances and takes negative values only close to the equilibrium substrate-adsorbate distance. Comparing the individual $\eta_{ij}$ contributions as obtained for the two considered adsorption geometries, one notes that while in each case the correlation contribution is more attractive when Xe is occupying the $atop$ position, the strongest deviation can be observed for $\eta_{12}$, $\eta_{14}$, and $\eta_{16}$ which relate to interaction between orbitals of the (S) and (S-1) Mg-layers (see Supplementary Fig. S4). This behavior can be explained in a following way: for the $atop$ geometry, the perturbation caused by Xe approaching the Mg surface is effectively screened by the metal (via redistribution of electron density); in the case of the $fcc$ geometry, such an effect does not take place. Similar conclusion can be done, when comparing the difference electron density plots as obtained by means of LDA calculations~\cite{scheffler}. 
 
\section{Conclusions}

Combining the incremental approach with the special embedding scheme, it is possible to consider metallic systems. While DFT with local and semi-local functionals can give reasonable results for cohesive energies, the errors are larger for surface stability and adsorption energies, and there is no systematic way of improving the accuracy. On the other hand, wavefunction-based correlation techniques, like CCSD(T), allow to get systematically improvable results for the full set of characteristics. Thus, the data obtained by means of presented approach can serve as  benchmarks in the development of improved DFT-based methods. Furthermore, the method of increments allows one not only to obtain reference values but also to understand influence of individual correlation-energy increments on the binding properties.

\section*{Acknowledgements}

The author would like to thank Prof. Dr. B. Pau\-lus (Berlin) and Prof. Dr. H. Stoll (Stuttgart) for valuable discussions. The computing facilities (ZEDAT) of the FU Berlin are acknowledged for computer time. The support from the German Research Foundation (DFG) through the Collaborative Research Center (SFB) 765 is appreciated. Worthful suggestions and useful discussions during preparation of the manuscript from Dr. Yu. S. Dedkov (Berlin) are acknowledged.

\clearpage
\begin{table*}
\caption{\label{tab:tab1} Lattice constants ($a$, $c$), cohesive energy ($E_\mathrm{coh}$), surface energy ($E_\mathrm{surf}$), equilibrium distance ($d_0$), and adsorption energy ($E_\mathrm{ads}$)  evaluated for the \textit{hcp} Mg, Mg(0001) surface, and Xe/Mg(0001) as computed for two adsorption geometries (\textit{c.f.} text), respectively, as obtained at different levels of theory. }
\begin{ruledtabular}
\begin{tabular}{lccc|c|cc|cc}
\multirow{3}{*}{Method}&\multicolumn{3}{c|}{\multirow{2}{*}{$hcp$ Mg}}                                    & \multirow{2}{*}{Mg(0001)}  & \multicolumn{4}{c}{ Xe/Mg(0001)} \\
              \cline{6-9}
              &    &&                                                                                   &                                                   & \multicolumn{2}{c|}{$atop$ position}& \multicolumn{2}{c}{$fcc$ hollow site}\\
              \cline{2-9}
              &$a$ (\AA)&$c$ (\AA)&$E_\mathrm{coh}$ (eV/atom)&$E_\mathrm{surf}$ (eV/atom)&$d_0$ (\AA)&$E_\mathrm{ads}$ (meV)&$d_0$ (\AA)&$E_\mathrm{ads}$ (meV)\\
\hline
Expt.                  &$3.21$$^a$&$5.21$$^a$&$ -1.51$$^a$&$+0.383$$^b$                                          & & &&\\
HF                     &$3.31$$^c$&$5.13$$^c$&$ -0.36$$^c$&$+0.182$                                                    &\multicolumn{2}{l|}{no binding}&\multicolumn{2}{l}{no binding}\\
CCSD(T)$^d$ &$3.21$$^e$&$5.14$$^e$&$-1.52$$^e$&$+0.385$                                                                      &$4.12$&$-72$&$4.28$&$-67$\\
LDA                  &$3.13$$^c$&$5.00$$^c$&$ -1.80$$^c$&$+0.404$                                                   &$3.83$&$-91$&$3.87$&$-85$\\
PBE                  &$3.19$$^c$&$5.12$$^c$&$-1.48$$^c$&$+0.360$                                                    &$4.81$&$-13$&$4.94$&$-12$\\
PBE-D2   &                      &                     &                       &                                                                     &$4.04$&$-75$&$4.13$&$-70$\\                 
\end{tabular}
\end{ruledtabular}
\raggedright
\footnotesize{$^a$Ref.~\cite{kittel}; $^b$Ref.~\cite{mg_surfen_expt}; $^c$Ref.~\cite{mg-wedig}; $^d$Correlation contribution to $E_\mathrm{coh}$, $E_\mathrm{surf}$, and $E_\mathrm{ads}$ are obtained according to eqns. (\ref{corr}), (\ref{surf}), and (\ref{ads}), respectively; $^e$Ref.~\cite{mg1}}
\end{table*} 

\clearpage
\begin{table}
\caption{\label{tab:tab2} Individual 2-body correlation energy increments (in eV) as obtained for ``bulk'' and ``surface'' clusters with CCSD(T) method. Correlation contributions to the surface energy ($\gamma$) are obtained when using eqn.~\ref{surf}. (Note: 1- and 3-body contributions are listed in Supplementary Tab. S1. Structure-related data for the considered increments are summarized in Supplementary Figs. S2 and S3).}
\begin{ruledtabular}
\begin{tabular}{l l |cc| cc| c}              
$ij$&$r_{ij}$&$\Delta\varepsilon_{ij}^\mathrm{bulk}$&$\omega_{ij}^\mathrm{bulk}$ &$\Delta\varepsilon_{ij}^\mathrm{surf}$& $\omega_{ij}^\mathrm{surf}$& $\gamma_{ij}$ \\
\hline
$12$&$3.197$&$-0.15411$&$ 3$&$-0.18050$&$3/2$&$+0.19158$\\
$13$$^a$&$3.209$&$-0.15190$&$ 3$&$-0.17156$&$   3$&$ -0.05898$\\
$14$&$4.529$&$-0.02768$&$ 3$&$-0.03299$&$3/2$&$+0.03355$\\
$15$&$5.210$&$-0.01375$&$ 1$&$-0.02217$&$1/2$&$+0.00267$\\
$16$&$5.551$&$-0.00982$&$ 6$&$-0.01272$&$   3$&$+0.02079$\\
$17$$^a$&$5.558$&$-0.00970$&$ 3$&$-0.01320$&$   3$&$ -0.01051$\\
$18$&$6.119$&$-0.00591$&$ 6$&$-0.00971$&$   3$&$+0.00633$\\
$19$$^a$&$6.418$&$-0.00515$&$ 3$&$-0.00730$&$   3$&$ -0.00645$\\
\end{tabular}
\end{ruledtabular}
\raggedright
\footnotesize{$^a$the both atoms of the central part of the cluster belong to the same layer of the $hcp$ lattice} 
\end{table}

\clearpage

\clearpage

\clearpage

\clearpage

\clearpage

\clearpage
\noindent
Supplementary material for the manuscript:\\
\textbf{Local correlation method for metals: benchmarks for surface and adsorption energies}\\
\newline
Elena Voloshina\\
\newline
\mbox{\textit{Physikalische und Theoretische Chemie, Freie Universit\"at Berlin, 14195 Berlin, Germany}}\\
\newline

\noindent \textbf{Content:}
\begin{itemize}

\item[1.]  Tab.\,S1:  Individual correlation-energy increments (in eV) as obtained for ``bulk'' and ``surface'' clusters with CCSD(T) method. 
\item[2.]  Fig.\,S1: The different contributions to the adsorption energy as functions of substrate-adsorbate distance are plotted for the Xe/Mg(0001) system as obtained for the two different adsorption geometries. Circles represent the data points, obtained by summing up the HF (thick solid line) and correlation (dashed line) parts of energy. Thin solid line is a fit according to the formula $E(d)=A\,\mathrm{exp}(-Bd)-C/(d-D)^3$ (\textit{c.f.} text).
\item[3.]  Fig.\,S2: The two-body correlation-energy increments are plotted as a function of distance.
\item[4.]  Fig.\,S3: The considered 3-body structures are shown schematically on the background of a hexagon in the $hcp$ plane.
\item[5.]  Fig.\,S4: Correlation contributions to the adsorption energy within the substrate as functions of the substrate-adsorbate distance, when considering the Xe/Mg(0001) system as obtained for (a) one-body increment $\eta_i$; (b-d) two body increments $\eta_{ij}$. (b): $\eta_{13}$, $\eta_{17}$, $\eta_{19}$ indicate interaction between orbitals within the (S) layer;  (c): $\eta_{12}$, $\eta_{14}$, $\eta_{16}$ indicate interaction between orbitals of the (S) and (S-1) layer; (d): $\eta_{15}$, $\eta_{18}$ indicate interaction between orbitals of the (S) and (S-2) layer. The energies are weighted for the total contribution of each increment. The scale is not the same in each subplot.

\end{itemize}

\linespread{1.2}
\clearpage

\begin{table*}
\raggedright
\noindent\textbf{Table\,S1.} Individual correlation-energy increments (in eV) as obtained for ``bulk'' and ``surface'' clusters with CCSD(T) method. Correlation contributions to the surface energy ($\gamma$) are determi- ned as a difference between the increment calculated using the ``surface'' cluster, multiplied with its weight factor ($\omega^\mathrm{surf}$), and the corresponding one calculated using the ``bulk'' cluster, multiplied with its weight factor ($\omega^\mathrm{bulk}$).\\

\begin{ruledtabular}
\begin{tabular}{l cc| cc| c}
                  &``bulk''                   &$\omega^\mathrm{bulk}$ &``surface'' &$\omega^\mathrm{surf}$   & $\gamma$ \\
\hline
\textbf{1}       &$-0.59990$&$   1$&$-0.65230$&$  1$&$ -0.05240$\\
$\sum$ 1-body       &$-0.59990$& &$-0.65230$& &$ -0.05240$\\
& & & &\\
\textbf{12}         &$-0.15411$&$ 3$&$-0.18050$&$3/2$&$+0.19158$\\
\textbf{13}$^a$&$-0.15190$&$ 3$&$-0.17156$&$   3$&$ -0.05898$\\
\textbf{14}         &$-0.02768$&$ 3$&$-0.03299$&$3/2$&$+0.03355$\\
\textbf{15}         &$-0.01375$&$ 1$&$-0.02217$&$1/2$&$+0.00267$\\
\textbf{16}         &$-0.00982$&$ 6$&$-0.01272$&$   3$&$+0.02079$\\
\textbf{17}$^a$&$-0.00970$&$ 3$&$-0.01320$&$   3$&$ -0.01051$\\
\textbf{18}         &$-0.00591$&$ 6$&$-0.00971$&$   3$&$+0.00633$\\
\textbf{19}$^a$&$-0.00515$&$ 3$&$-0.00730$&$   3$&$ -0.00645$\\
$\sum$ 2-body              &$-1.15375$&        &$-0.97479$&          &$+0.17896$\\
& & & &\\
\textbf{3.1}[2+1]$^b$&$ -0.00276$& $3$&$+0.00785$&$5/2$&$+0.02791$\\
\textbf{3.1}[1+2]$^c$&$ -0.00276$& $3$&$+0.01637$&$   2$&$+0.04102$\\
\textbf{3.2}$^a$  &$+0.00198$& $1$&$+0.00534$&$   1$&$+0.00336$\\
\textbf{3.3}$^a$  &$ -0.00596$& $1$&$ -0.00272$&$   1$&$+0.00324$\\
\textbf{3.4}[2+1]$^b$&$+0.00078$& $6$&$+0.00173$&$   5$&$+0.00397$\\
\textbf{3.4}[1+2]$^c$&$+0.00078$& $6$&$+0.00756$&$   4$&$+0.02556$\\
\textbf{3.5  }          &$ -0.00134$& $3$&$ -0.00324$&$5/2$&$ -0.00408$\\
\textbf{3.6}[2+1]$^b$&$ -0.00132$& $6$&$ -0.00498$&$   5$&$ -0.01698$\\
\textbf{3.6}[1+2]$^c$&$ -0.00132$& $6$&$+0.00076$&$   4$&$+0.01096$\\
\textbf{3.7}$^a$ &$ -0.00130$& $6$&$ -0.00163$&$   6$&$ -0.00198$\\
\textbf{3.8}          &$ -0.00244$& $6$&$ -0.00582$&$   5$&$ -0.01446$\\
\textbf{3.9}$^a$ &$ -0.00271$& $3$&$ -0.00320$&$   3$&$ -0.00147$\\
$\sum$ 3-body               &$ -0.06161$&         &$+0.01544$&          &$+0.07705$\\
& & & &\\
$E^{corr}_\mathrm{total}$
                             &$ -1.81526$ &         &$ -1.61165$&          &$+0.20361$\\
\end{tabular}

\end{ruledtabular}
\raggedright
\footnotesize{$^a$all atoms of the central part of the cluster belong to the same layer (S) of the $hcp$ lattice;\\
$^b$two atoms of the central part belong to the (S)-plane and the remaining atom is lying in the (S-1) plane;\\
$^c$one atom of the central part belongs to the (S)-plane and the remaining two atoms are lying in the (S-1) plane
} 
\end{table*}

\clearpage
\begin{figure}
\includegraphics[width=1.0\textwidth]{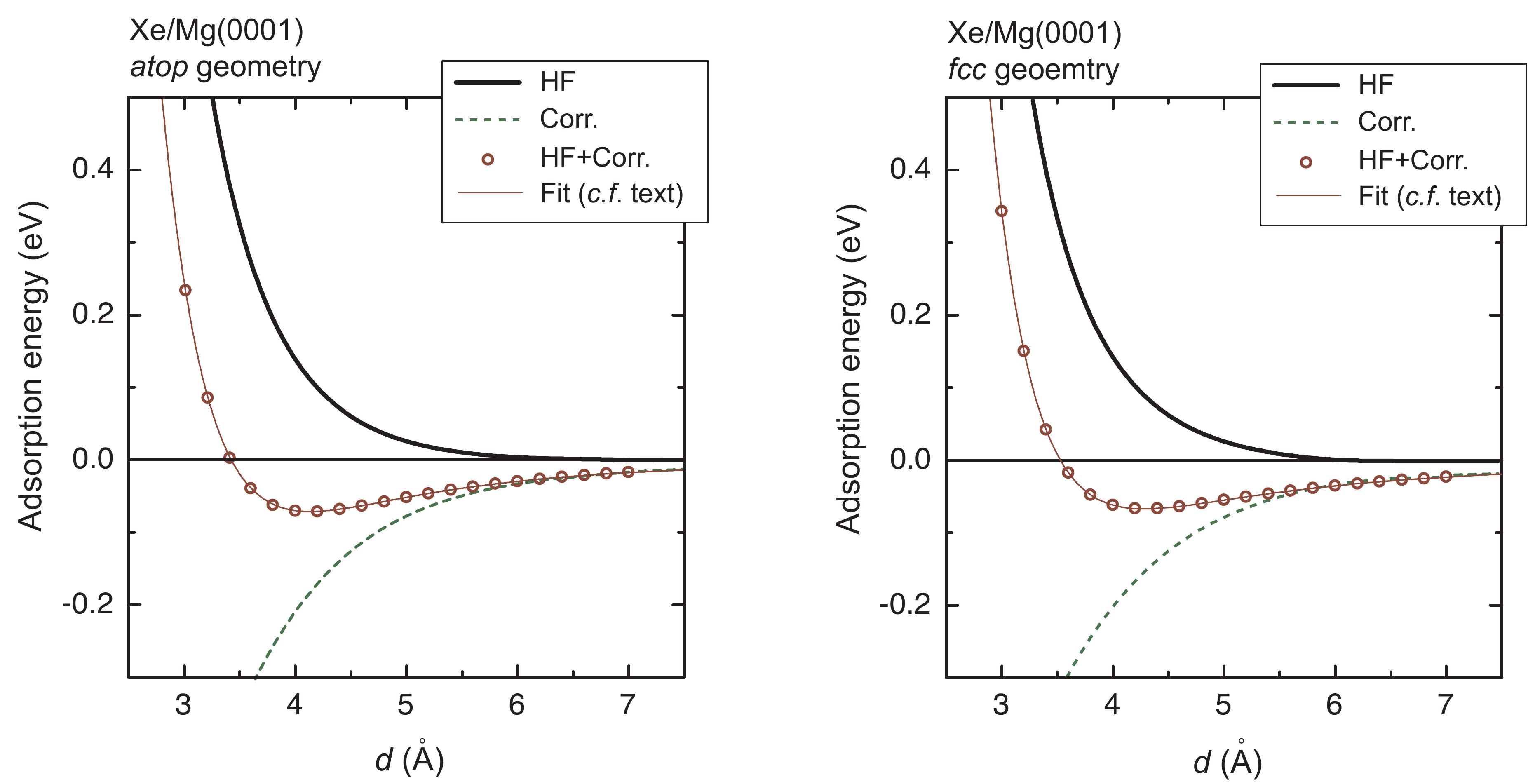}
\end{figure}
\noindent\textbf{Fig.\,S1.} The different contributions to the adsorption energy as functions of substrate-adsorbate distance are plotted for the Xe/Mg(0001) system as obtained for the two different adsorption geometries. Circles represent the data points, obtained by summing up the HF (thick solid line) and correlation (dashed line) parts of energy. Thin solid line is a fit according to the formula $E(d)=A\,\mathrm{exp}(-Bd)-C/(d-D)^3$ (\textit{c.f.} text).

\clearpage
\begin{figure}
\includegraphics[width=0.65\textwidth]{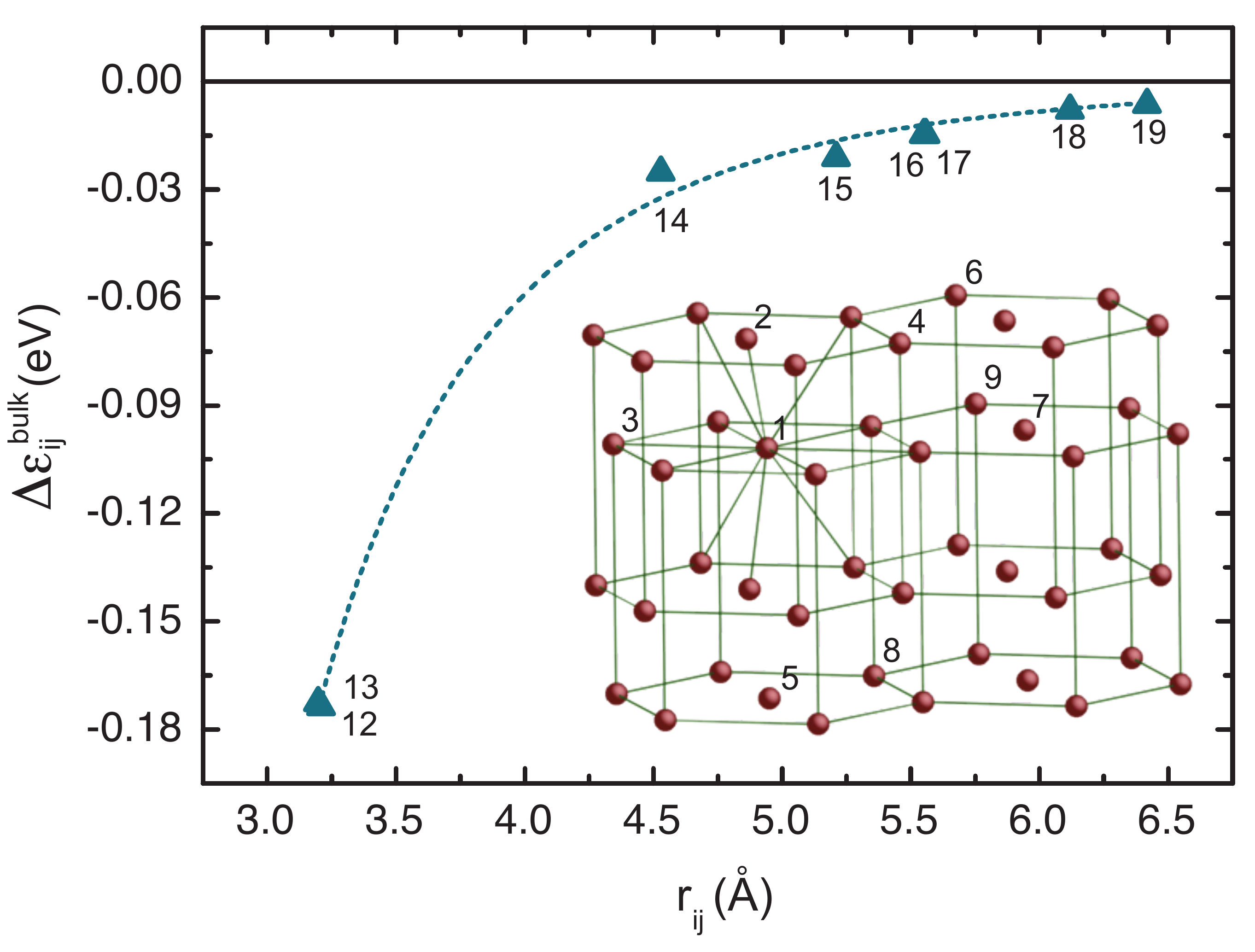}
\end{figure}
\noindent\textbf{Fig.\,S2.} The two-body correlation-energy increments are plotted as a function of distance. Inset: In the crystallographic structure of $hcp$ Mg, the atoms are ordered by their distance from the atom number 1. 

\clearpage
\begin{figure}
\includegraphics[width=0.65\textwidth]{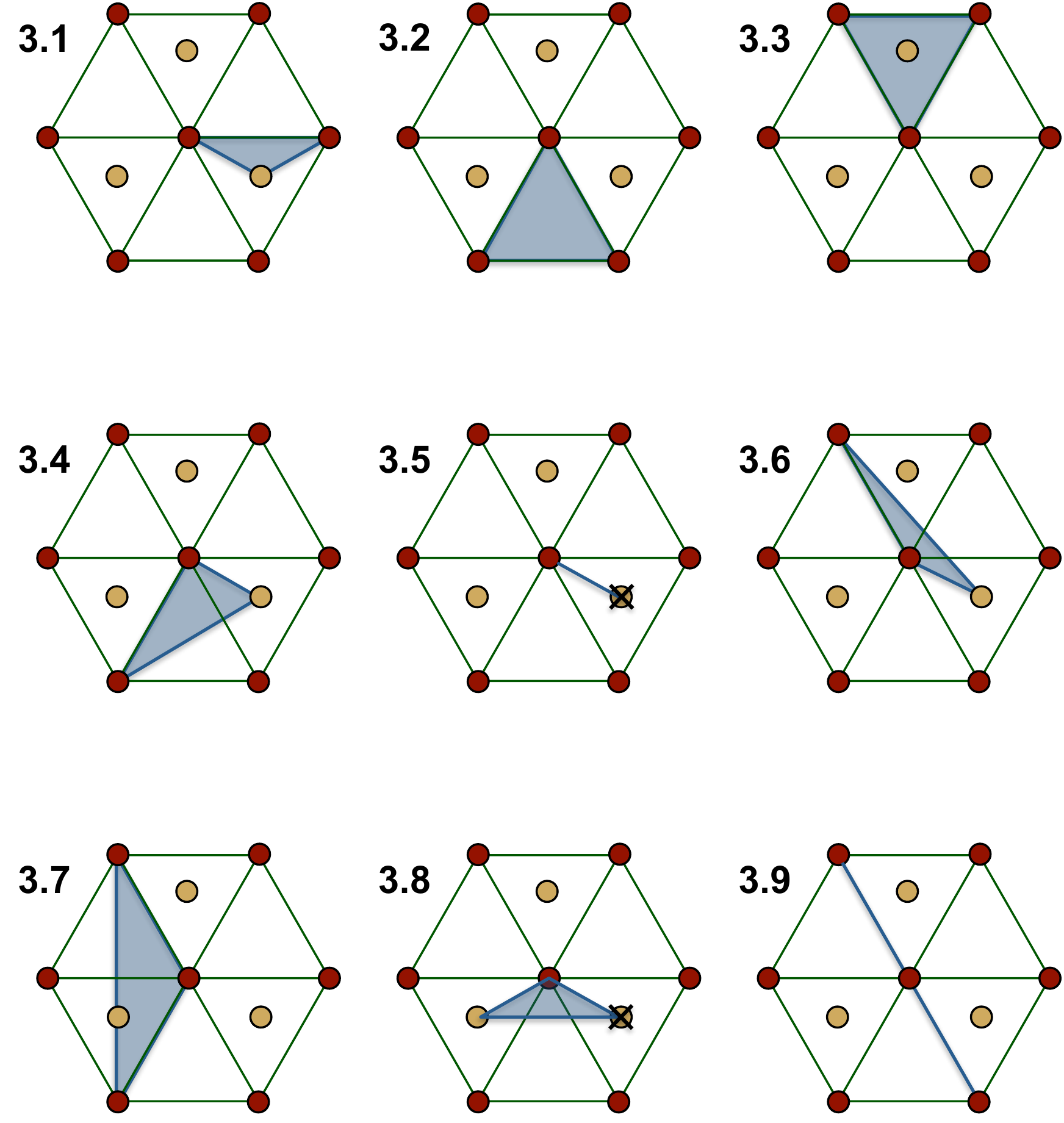}
\end{figure}
\noindent\textbf{Fig.\,S3.} The considered 3-body structures are shown schematically on the background of a hexagon in the $hcp$ plane. Light spheres correspond to atoms below the plane, and crosses to atoms above the plane. 

\clearpage
\begin{figure}
\includegraphics[width=1.0\textwidth]{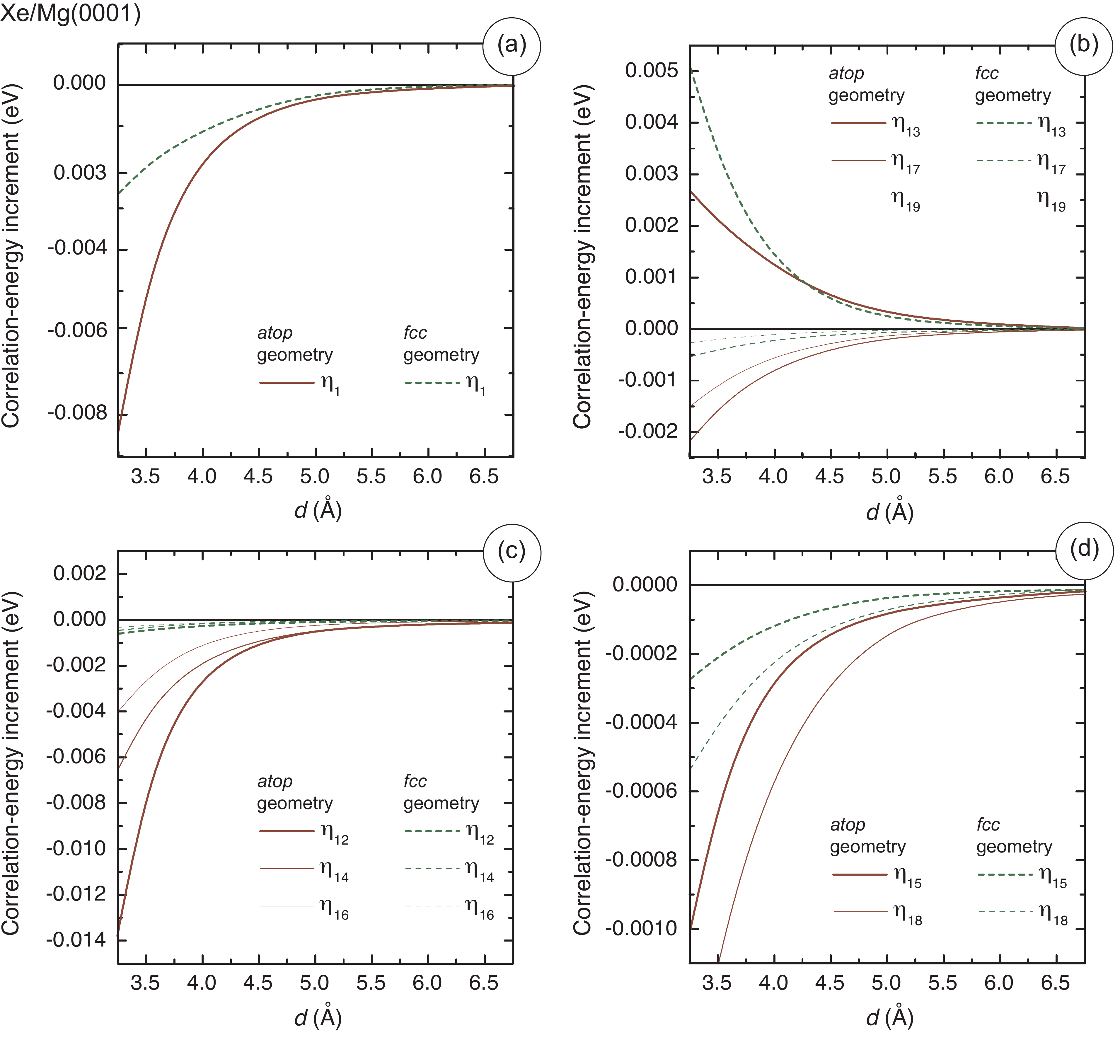}
\end{figure}
\noindent\textbf{Fig.\,S4.} Correlation contributions to the adsorption energy within the substrate as functions of the substrate-adsorbate distance, when considering the Xe/Mg(0001) system as obtained for (a) one-body increment $\eta_i$; (b-d) two body increments $\eta_{ij}$. (b): $\eta_{13}$, $\eta_{17}$, $\eta_{19}$ indicate interaction between orbitals within the (S) layer;  (c): $\eta_{12}$, $\eta_{14}$, $\eta_{16}$ indicate interaction between orbitals of the (S) and (S-1) layer; (d): $\eta_{15}$, $\eta_{18}$ indicate interaction between orbitals of the (S) and (S-2) layer. The energies are weighted for the total contribution of each increment. The scale is not the same in each subplot.

\end{document}